\documentclass[10pt,conference]{IEEEtran} 
\IEEEoverridecommandlockouts
\usepackage{float}
\usepackage{cite}
\usepackage{graphicx}
\usepackage{listings}
\usepackage[inline]{enumitem}
\usepackage[left=1.62cm,right=1.62cm,top=1.96cm]{geometry}
\usepackage{algorithm}
\usepackage{algpseudocode}
\usepackage{varwidth}

\begin{document}
\title{Access control for Data Spaces}

\author{
\IEEEauthorblockN{
Nikos Fotiou,\IEEEauthorrefmark{1}
Vasilios A. Siris,\IEEEauthorrefmark{1}\IEEEauthorrefmark{2}
George~C.~Polyzos\IEEEauthorrefmark{1}\IEEEauthorrefmark{2}\IEEEauthorrefmark{3}
}
\\
\IEEEauthorblockA{\IEEEauthorrefmark{1}
ExcID P.C., 11362 Athens, Greece}
\IEEEauthorblockA{\IEEEauthorrefmark{2}
Department of Informatics, School of Information Sciences and Technology, \\
Athens University of Economics and Business, 10434 Athens, Greece}
\IEEEauthorblockA{\IEEEauthorrefmark{3}
School of Data Science, The Chinese University of Hong Kong, Shenzhen, 518172 Guangdong, China}

\vspace{-0.3in}
}

\maketitle

\begin{abstract}
Data spaces represent an emerging paradigm that facilitates secure and trusted data exchange through foundational elements of data interoperability, sovereignty, and trust. Within a data space, data items, potentially owned by different entities, can be interconnected. Concurrently, data consumers can execute advanced data lookup operations and subscribe to data-driven events. Achieving fine-grained access control without compromising functionality presents a significant challenge. In this paper, we design and implement an access control mechanism that ensures continuous evaluation of access control policies, is data semantics aware, and supports subscriptions to data events. We present a construction where access control policies are stored in a centralized location, which we extend to allow data owners to maintain their own Policy Administration Points. This extension builds upon %two authorization methods: one based on OAuth~2.0 and another utilizing 
W3C Verifiable Credentials.  
%We evaluate both authorization methods and discuss their trade-offs, concluding that there is no one-size-fits-all solution as each method offers distinct security, privacy, and availability characteristics.    
\end{abstract}


\section{Introduction}\label{introduction}

Data spaces are emerging as a new form of digital platform aiming at 
% deliberating
% liberating data from 
eliminating
silos
% in order to 
and
enabling data-driven innovations and shape the digital transformation~\cite{Bev2022}. 
A growing number of reports by commercial entities and governmental bodies highlight the business potential and the possible societal impact that can be achieved  by embracing data spaces (see for example~\cite{Sce2022}). A data space is composed of building blocks that enable semantic interoperability of data, uniform data access methods, as well as increased data sovereignty and trust. Nevertheless, data usage policies and access control still remain a challenge for data spaces~\cite{Gel2020, Sch2022}. Data owners do not want to lose control and sovereignty over their data;
% similarly, 
data users want to safeguard against low-quality or malicious data~\cite{Reu2022}. %Decentralization of identity and of trust mechanisms can significantly contribute towards this direction, however this requires~\cite{Ibr2023}: a holistic Identity and Access Management system, decentralized management of organizational credentials, self-sovereign management of user credentials, policy foundation and trust services, as well as decoupling trust frameworks. 

%subsection{An IoT data space}
%\begin{figure}
%  \includegraphics[width=0.9\linewidth]{data-space-crop.pdf}
%  \caption {Data space overview.}
%  \label{fig:dspace}
%\end{figure}
%In the remainder of this paper, 
We use a data space for exchanging data generated by IoT devices as a motivating use case. This data space  consists of IoT device \emph{owners} who wish to share data generated by their devices, acting as data \emph{suppliers}, data \emph{consumers} who seek to access the provided data through a \emph{client application}, and data \emph{intermediaries} that offer a data \emph{context broker} that implements the corresponding data management and access APIs.  
% and \emph{trust providers} that govern trust relationships.

A context broker 
maintains digital objects,
%is used to maintain digital objects, 
which in our use case are \emph{digital twins} of the corresponding IoT devices. Each such object is uniquely identifiable and serialized using JSON for \emph{Linked Data} (JSON-LD)~\cite{jsonld}. An object has a \emph{type} and \emph{attributes}. Accordingly, a type can be associated with many objects and an object may have multiple attributes. For example, a data space may include objects of type \emph{smart lamp}, with identities such as \emph{lamp1} and \emph{lamp2}, and attributes such as \emph{consumption}, \emph{status}, and \emph{color}.

A context broker implements 
%the \emph{Next Generation Service Interfaces Linked Data} (NGSI-LD) API~\cite{ngsild}, which is an ETSI standard  that allows HTTP-based operations 
the ETSI standard \emph{Next Generation Service Interfaces Linked Data} (NGSI-LD) API~\cite{ngsild}, which allows HTTP-based operations
to digital objects stored in the broker, as well as subscriptions to events related to these objects. This API simultaneously allows fine-grained data access, e.g., a consumer may request only specific attributes of an object, as well as coarse-grained access, e.g., a consumer may request attributes of all objects of a specific type.

\subsection{Access control}
A context broker should be protected by an access control mechanism that allows access to stored data  only by authorized data consumers. An access control solution should achieve:
% the following:

\begin{itemize}
\item \textbf{Attack surface reduction:} An access control solution should strive to minimize potential security threats. The amount of security-sensitive information managed by data consumers should be minimal. Similarly, access verification should be simple and not prone to errors. %Finally, the principle of least privilege should be enforced.
\item 
\textbf{Usage control:} Access rights of a consumer should be re-evaluated, even after a consumer has been initially authorized by a data intermediary. For example, a consumer that has successfully subscribed to receive notifications about an object should stop receiving new notifications if its corresponding access rights are revoked. 
\item 
\textbf{Enhanced privacy:} An access control solution should prevent tracking of data consumers not only by third parties, but also by participants of the data space (e.g., by data intermediaries and owners).
\item 
\textbf{Availability:}  An access control solution should not  depend on (Internet) connectivity, instead it should function correctly  even if some of its components are unreachable.   
\end{itemize}

\subsection{Contributions}
In this paper, we present the design and evaluation of an access control solution tailored to data spaces, making the following key contributions: (i) we introduce access control policies that are aware of data semantics (e.g., a consumer can be authorized to access all objects of a certain type); (ii) we enable continuous monitoring and re-evaluation of consumer access rights, supporting long-term operations such as event subscriptions; (iii) we facilitate distributed deployments, allowing each data owner to maintain their own \emph{Policy Administration Point}, thereby enhancing owner sovereignty. %Our design leverages two prevalent standards: OAuth 2.0 and W3C Verifiable Credentials.

In the remainder of this paper, we detail our % solution's 
design in Section II, 
% we 
present its 
% design 
implementation and 
% security 
evaluation in Section III, and 
% we conclude our paper 
our conclusions and future work
in Section IV.

\section{Design}

\subsection{Underlay Data Space}
%\begin{figure*}
%  \includegraphics[width=0.9\linewidth]{objects-crop.pdf}
%  \caption {Data space object organization and data object requests. Orange circles represent consumer requests and green circles represent broker responses. 
  %Orange circles represent what the consumer requested and green circles represent the broker response. 
%  }
%  \label{fig:obj}
%\end{figure*}
In the considered data space, data owners create new objects and assign policies specifying the operations a data consumer can perform on an object. Objects and types are uniquely identified using a URL denoted by $URL_{object}$ and $URL_{type}$, respectively.
Similarly, object attributes are identified by a $URL_{attr}$, which has the form $URL_{object}/attribute name$. Consumers are also identified by an identifier denoted by $Consumer_{Id}$. %The identifier of a consumer is associated with a public/private key pair that the consumer can use to prove identity possession. 
Consumers can request access to all objects of a certain type, or to specific objects. 
%(a) specific object(s) (right part of Fig.~\ref{fig:obj}). 
% overview. 
Additionally, consumers can provide filters specifying which attributes they wish to access. Consumers can perform read, write, and subscription operations over stored objects using the ETSI NGSI-LD API and the corresponding endpoint of the context broker.   

\begin{figure}[H]
\vspace{-0.5cm}
\centering
  \includegraphics[width=0.75\linewidth]{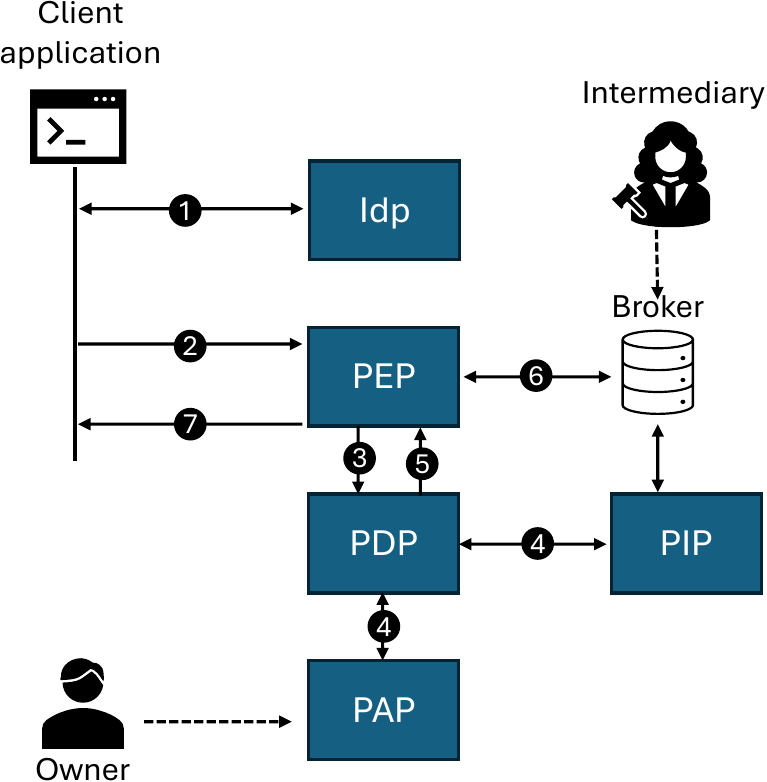}
\vspace{-0.2cm}
  \caption {High level overview of the authorization process.}
  \label{fig:authz}
\vspace{-0.25cm}
\end{figure}

\subsection{Components and interactions}

Our access control solution is composed of the following components:

\paragraph{Identity Provider (IdP)} A registry of consumer identifiers. This registry is maintained by a party trusted by owners, intermediaries, and consumers; this can be a Trusted Third Party or any of these three entities.

\paragraph{Policy Administration Point (PAP)} A registry where owners can store access control policies that define the access rights of each consumer. 

\paragraph{Policy Enforcement Point (PEP)}  A transparent HTTP proxy that intercepts API calls from consumers to the context broker. If a request originates from a consumer with the appropriate access rights it is forwarded to the (context) broker, otherwise it is rejected. 

\paragraph{Policy Decision Point (PDP)}  A component that decides whether or not a request originates from an authorized consumer. 

\paragraph{Policy Information Point (PIP)} A component that provides supplementary information used by a PDP to make an access control decision. Specifically, this component provides details about the attributes and type of an object.

From a high-level perspective, authorization using our solution is implemented as follows (see also Fig.~\ref{fig:authz}). The IdP is configured with consumer identifiers, e.g., through a client registration process. Similarly, an owner configures the PAP with the appropriate access control policies. A consumer initially, through its client application, identifies itself to the IdP and receives an \emph{identity token} (step 1), i.e., a token signed by the IdP that includes a proof of the consumer's identifier. Then it makes an NGSI-LD API call, including the received token in an HTTP header (step 2).  The request is intercepted by the PEP, which forwards it to the PDP that makes 
% an 
the
access control decision (step 3). The PDP collects the consumer's access rights  from the PAP and, if necessary, information related to the requested object from the PIP (step 4). Using the retrieved  information, the PDP makes an access control decision 
% which 
that it forwards to the PEP (step 5).
Finally, if the consumer is authorized, the PEP forwards the request to the context broker (step 6) and relays its response back to the consumer (step 7).

\subsection{Policies and enforcement}
Our system implements capabilities-based access control, where  
%consequently, 
an access control policy specifies the operations (i.e., Read, Write, Subscribe) that a consumer can perform on specific object types, and/or on specific objects, and/or on specific object attributes. For example, a consumer can be allowed to \emph{Read all objects of type ``smart lamp'', and Write the attribute ``energy consumption'' of the objects with identifiers ``smart lamp1'' and ``smart lamp2''}. More formally, we define an access control policy $p$ as the tuple $[Consumer_{id}, operation, URL]$, where $operation$ can be Read, Write, or Subscribe, and $URL$ can be a $URL_{type}$ or $URL_{object}$ or $URL_{attr}$. We define an access control decision as the following function:
$$Decide(Consumer_{id}, operation, URL)\rightarrow\{true,false\}$$
In order to enable access control decisions we define the $URL_A \supseteq URL_B$ operation which outputs $true$ if one of the following conditions hold:
\begin{itemize}
    \item $URL_A = URL_B$
    \item $URL_A$ is a $URL_{type}$ and $URL_B$ a $URL_{object}$ of an object of type $URL_A$ 
    \item $URL_A$ is a $URL_{type}$ and $URL_B$ a $URL_{attr}$ of an attribute of an object of type $URL_A$
    \item $URL_A$ is a $URL_{object}$ and $URL_B$ a $URL_{attr}$ of an attribute of object $URL_A$
\end{itemize}

%For example, consider the data space of Fig.~\ref{fig:obj}, the following conditions hold:
%$$URL_T \supseteq URL_T = true$$
%$$URL_T \supseteq URL_{O1} = true$$
%$$URL_T \supseteq URL_{O1}/A1 = true$$
%$$URL_{O1} \supseteq URL_{O1}/A1 = true$$
%$$URL_{O1} \supseteq URL_{O2}/A1 = false$$
%a type $/lamps$ and an object $/lamp1$ of type $/lamps$ with an attribute $/lamp1/status$. The following operations output $true$: $/lamps \supseteq /lamps$, $/lamps \supseteq /lamp1$, $/lamps \supseteq /lamp1/status$, $/lamp1 \supseteq /lamp1/status$
The access control decision is implemented using Algorithm~\ref{alg:cap}. 
% The 
Its
semantics 
% of this algorithm is 
denote
that if a consumer is authorized to perform an operation on a type, it can perform the same operation on all objects of this type and their attributes. Similarly, if a consumer is authorized to perform an operation on an object, it is authorized to perform the same operation on all its attributes.

\begin{algorithm}
\caption{Access control decision algorithm}\label{alg:cap}
\begin{algorithmic}
\State Let $P$ the set of all policies
\Procedure{Decide}{$Consumer_{id},operation,URL$}
\ForAll{$p \in P$}
\If{$p[Consumer_{id}] = Consumer_{id}$ AND \\ 
\hskip\algorithmicindent\hskip\algorithmicindent\hskip\algorithmicindent  $p[operation] = operation$ AND \\ \hskip\algorithmicindent\hskip\algorithmicindent\hskip\algorithmicindent $p[URL] \supseteq URL$}
\State \textbf{return} $true$
\EndIf
\EndFor
\State \textbf{return} $false$
\EndProcedure
\end{algorithmic}
\end{algorithm}
% Additionally, in 
In many cases, 
% in order 
for the PDP to make an access control decision the type of the requested object must be known: this \emph{type inference} functionality is provided by the $PIP$. For example, assume that
%let a 
consumer $C$ has received authorization for the \emph{Read} operation on object type $T$ (e.g., authorization to \emph{Read all objects of type ``smart lamp''}). $C$ makes a \emph{Read} request for attribute $A$ of object $O_1$ (e.g.,  a request to \emph{Read the ``status'' of ``smart lamp 1''}).  For the PDP to make an access control decision it needs to infer the type of $O_1$. This is achieved using the PIP,  which communicates with the context broker and obtains
%learns 
the type of $O_1$ using the corresponding NGSI-LD API call. Then, the PDP uses this information to make a decision, i.e., if $O_1$ is of type $T$ the request is accepted.

Finally, our solution provides \emph{Automatic un-subscription}. 
Specifically, 
%Particularly, a
the PDP maintains a list of active subscriptions and automatically un-subscribes
% (by invoking the corresponding API call) 
(invoking an API call) 
% a 
consumers
% when 
that
% the consumer 
% is 
are
no longer authorized to receive 
% a 
notifications (e.g., the access rights have expired or were revoked).

\subsection{Distributed PAPs}
We now present an extension 
% to our solution 
allowing PAPs to be managed by the corresponding owners
%. This extension 
and
provides increased security and improved sovereignty for data owners. We leverage W3C Verifiable Credentials (VCs) to enable intermediary's access control components to learn the capabilities of a consumer.

A 
% Verifiable Credential (VC) 
VC
is a W3C recommendation that allows an \emph{issuer} to assert some claims about an entity, referred to as the VC \emph{subject}.
A VC includes information about the issuer, the subject, the asserted claims, as well as possible constraints (e.g., expiration date)~\cite{ver}. This information is encoded in a machine readable format (e.g., as a JSON object in our system). Then, a VC holder (usually, the VC subject itself) can prove to a verifier that it owns one or more VCs with certain characteristics. This is achieved by binding VCs to a subject identifier (e.g., a public key) that can be used for generating a \emph{Verifiable Presentation} (VP) of the VC(s). A VP is an object (a JSON object in our system) that includes one or more VCs and it is signed in a way that can be verified using the subject identifier (specified in the included VC(s)). VP verification does not require communication with the issuer. 

\begin{figure}[H]
\vspace{-0.45cm}
\centering
  \includegraphics[width=0.9\linewidth]{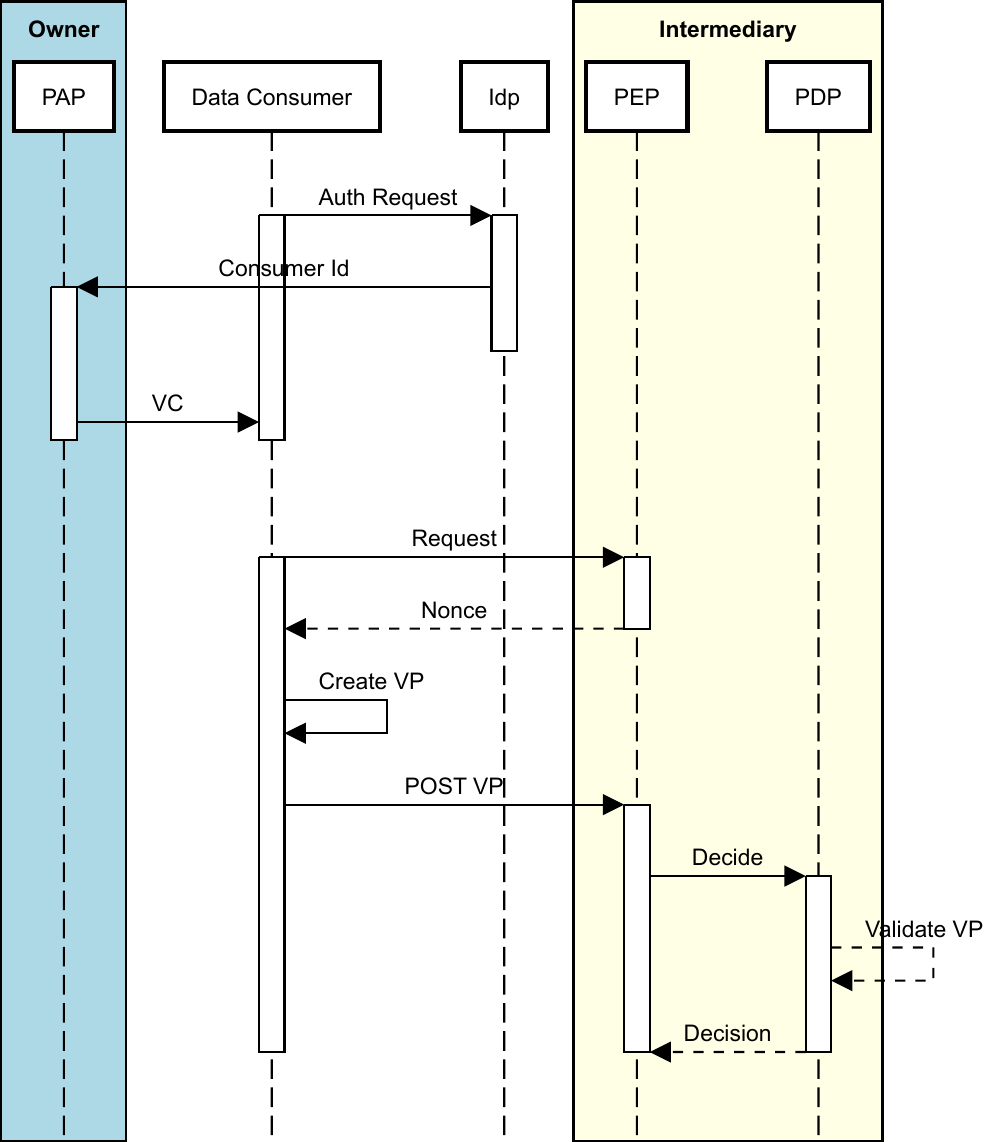}
  \caption {Authorization using Verifiable Credentials.}
  \label{fig:vc}
\vspace{-0.25cm}
\end{figure}

In our data space solution, a VC can be used as a means for communicating consumer capabilities (see also Fig.~\ref{fig:vc}). In this case, a VC is issued by a PAP to a consumer and  is serialized as a JSON object that includes the consumer access rights and a public key provided by the consumer; this JSON object is digitally signed by the PAP. VC issuance is implemented using OpenID for Verifiable Credential Issuance~\cite{oid4vci}. Using this protocol a consumer authenticates to the PAP (acting as the VC \emph{issuer}) using the IdP and OpenID connect. Then, the PAP issues the corresponding credential. 

Data consumer authorization is implemented using OpenID for Verifiable Presentation~\cite{oid4vp} and  involves the following steps: First, a consumer sends an \emph{unauthorized} request and receives a \emph{nonce}. Then, the consumer creates and submits a Verifiable Presentation (VP). The VP includes a valid VC, the received nonce, and the HTTP URL of the intermediary, and  is signed using the private key of the consumer identifier. Then, the PDP endpoint performs the following validations:
\begin{itemize}
  \item It verifies that the VC has been issued by a trusted PAP.
  \item It verifies the correctness and the validity of the VC by validating its digital signature, its issuance and expiration time, and by querying for its revocation status (see the following section.)
  \item It validates the VP proof using the consumer's public key included in the VC, and  verifies that it includes a valid nonce and intermediary URL.  
\end{itemize}

\subsection{Usage control}
Consumer capabilities are serialized using VCs 
% that includes an \emph{expiration} time.
that include an \emph{expiration} time.
Capabilities can be cached by the PEP up until their expiration time for two reasons: (i) to avoid consumer re-authorization, and (ii) to enable access control on subscriptions.
%However, while they are cached, capabilities may be revoked. For this reason a PDP needs to be able to verify their validity periodically, which is challenging when PAPs are distributed.
However, while they are cached, capabilities may be revoked, thus a PDP needs to be able to verify their validity periodically, which is challenging when PAPs are distributed.  
%In our OAuth~2.0 approach we leverage \emph{refresh tokens}. Specifically,  with the token exchange process a PDP  also receives a \emph{refresh token}. This token can be used periodically by the PDP to request from the PAP a new security token that provides an ``update'' of the consumer's capabilities. The advantage of this approach is that it can be used not only for examining if a consumer's capabilities have been revoked, but also if new capabilities have been added. Additionally, this approach allows continuous fine-grained capabilities management, e.g., it allows owners to revoke only a subset of the consumer's capabilities. 
% For this purpose our solution supports VC revocation. 
For this reason our solution supports VC revocation. 
Specifically, we rely on a recent W3C draft~\cite{revoclist} that defines an efficient revocation mechanism.
%In more detail, to 
To support revocation a PAP maintains a revocation list that covers all \emph{non-expired} VCs that it has issued. This list is a simple bit string and each VC is associated with a position in the list. Consequently, each VC includes a property named \emph{credentialStatus} that specifies the position of that VC in the revocation list, as well as a URL that can be used for retrieving the revocation list. A VC is simply revoked by setting the corresponding bit in the revocation list to $1$. Since the list includes only non-expired VCs, its size is tolerable for most use cases. Similarly, since this list is expected to include many $0$s and few $1$s it can be efficiently compressed. A revocation list is included in a VC, issued by the PAP, which is periodically downloaded by the PDP. Nevertheless, since a revocation list covers all non-expired VCs issued by the PAP, the frequency at which a PDP downloads this list does not depend on the number of stored capabilities. On the other hand, this approach does not allow capability updates, neither fine-grained capabilities-management: once a VC is revoked, all consumer capabilities stored by the PDP are removed. 

\section{Implementation}
\label{implementation}
We implemented a prototype of our data space
%, which is used 
% for sharing data collected by IoT devices installed in real smart buildings. 
for sharing smart building IoT device data. 
% In a nutshell, data items collected by the IoT devices are stored in the back-end of two companies (acting as the data suppliers). There, they are serialized using JSON-LD, and stored in a context broker of a data intermediary implemented using FIWARE Orion~\cite{Ahl2022}. 
Data are stored in the back-ends of two companies (acting as the data suppliers).
They are serialized using JSON-LD and stored in a context broker 
% of a data intermediary 
implemented using FIWARE Orion~\cite{Ahl2022}. 

%The communication between a data supplier, a data consumer, the PAP, and  the data intermediary takes place over HTTPS and it is assumed to be secured. 

As an IdP we used keycloack.\footnote{https://www.keycloak.org/}
A PAP has also been implemented following the iSHARE trust framework\footnote{https://ishare.eu/}. Using this framework, owners can specify access control policies that define the access rights of each data consumer using the iSHARE policy definition language,\footnote{https://dev.ishare.eu/delegation/policy-sets.html} which is inspired by XACML. 
% As described previously, a
A policy defines the operations that a consumer
% , identified by a specific identifier, 
can perform over a set of object types, and/or a set of object identifiers, and/or a set of object attributes. The corresponding PDP and PIP have been implemented as custom applications.
Finally, for the  PEP we rely on Ory Oathkeeper.\footnote{https://www.ory.sh/oathkeeper/}

%Then, the consumer communicates with the context broker. All communication between the consumer and the context broker is intercepted by a \emph{policy enforcement point} (PEP). The PEP executes an authorization protocol for requests of unauthorized consumers and forwards requests of authorized consumers to the context broker. Information about the participants of the data space is stored in a \emph{satellite} service provided by iSHARE. 

%Data consumers and associated with a public key which is included in a digital certificate, issued and signed by a trusted certificate authority referred to as the \emph{satellite}. Our implementation relies on iSHARE satellite service.\footnote{https://ishare.eu/ishare-satellite-explained/} In the OAuth~2.0-based approach, data consumers' public key is used for authenticating to the IdP, whereas in our VC-based approach this key is additionally used for signing a VP.  The satellite also provides an API endpoint that allows $3^{rd}$ parties to request information about an entity. Particularly, given an entity identifier, this endpoint responds with the corresponding certificate and \emph{status}; the status can be either \emph{active} or \emph{inactive} (e.g., because the certificate has been revoked). 

\subsection{Security evaluation}
We now evaluate the security properties of our solution focusing on the distributed version by revisiting the security requirements defined in Section~1. Our solution makes the following security assumptions. The communication of all entities in our system takes place over secure communication channels; similarly, all cryptographic operations are secure. Furthermore, it is assumed that IdPs and PAPs are trusted, as well as legitimate intermediaries operate as expected. Finally, it is assumed that all secrets are properly secured. 

\subsubsection{Attack surface reduction}
%OAuth~2.0 is a widely used standard with many, well-supported and verified implementations. In our OAuth~2.0 approach a data consumer needs to protect a secret used for authenticating to the IdP, as well as the public/private key pair used for signing the DPoP proof. Similarly, and according to~\cite{rfc8693}, an intermediary needs to maintain an authentication method for each PAP it interacts with in order to receive a service token. This can be a single public/private key pair or many pre-shared secrets. Service tokens are received directly from the PAP, therefore, a PDP does not have to make any additional verification checks. 
%Similarly, and in order to comply with OAuth~2.0 specifications, a Token endpoint needs to have established an authentication mechanism with the PAP (in order to receive the token); in our system, a sufficiently long secret key is used for this purpose. Since the Token endpoint interacts with the PAP directly over HTTPS, it does not need to perform any validation on the received evidence.   

The W3C VC data model is a W3C recommendation, for which there are not many implementations available. In our VC-based approach, a data consumer needs to protect a secret used for authenticating to the IdP, as well as 
% public/private key pair 
the private key 
used for signing VPs. Since a VC is received from a data consumer (and not from the PAP directly), a PDP needs to verify its integrity by validating its signature, which has been generated by the PAP. 

Our threat model considers malicious entities trying to gain unauthorized access to a data space. These entities operate as (malicious) intermediaries and their goal is to gain access to a data space provided by another (legitimate) intermediary. More formally, let $Auth(C,I,o) \rightarrow true$ if consumer $C$ is authorized to access an object $o$ in a data space provided by intermediary $I$. Additionally, let $I_{leg}$, $I_{mal}$, $C_1$, and $o_1$ be a legitimate intermediary, a malicious intermediary, a consumer, and an object respectively. Supposedly, 
$$Auth(C,I_{leg},o) \rightarrow true ~~~~~
%\space\space\space
%$$ 
%$$
Auth(C,I_{mal},o) \rightarrow true$$
The goal of $I_{mal}$ is to achieve 
% $Auth(I_{leg},I_{leg},o) \rightarrow true$. 
$Auth(I_{mal},I_{leg},o) \rightarrow true$. 
Due to the second assumption, $I_{mal}$ has access to a VP  generated by $C$. In order to achieve its goals it should be able to use them in a request towards $I_{leg}$. However, a VP includes the URL of $I_{mal}$ and since $I_{mal}$ cannot modify them, they will be rejected by $I_{leg}$.

\subsubsection{Usage Control}
\label{sec:rev}
Usage control is implemented using revocation lists. In this case some overhead is introduced.

\paragraph{Initial authorization overhead}  VC issuance and authorization using a VP are asynchronous processes, therefore the PDP must verify the revocation status of a VC immediately after a VP has been received. This introduces
% some 
additional delay the first time a consumer tries to interact with the data space. 

\paragraph{Authorization modification overhead} In case the capabilities of a consumer are modified, old consumer capabilities should be revoked in order for the PDP to request a new VC from the consumer. Alternatively, a consumer can pro-actively ``refresh'' its VCs; this
% method 
is left as future work. 

\begin{figure}
  \includegraphics[width=0.9\linewidth]{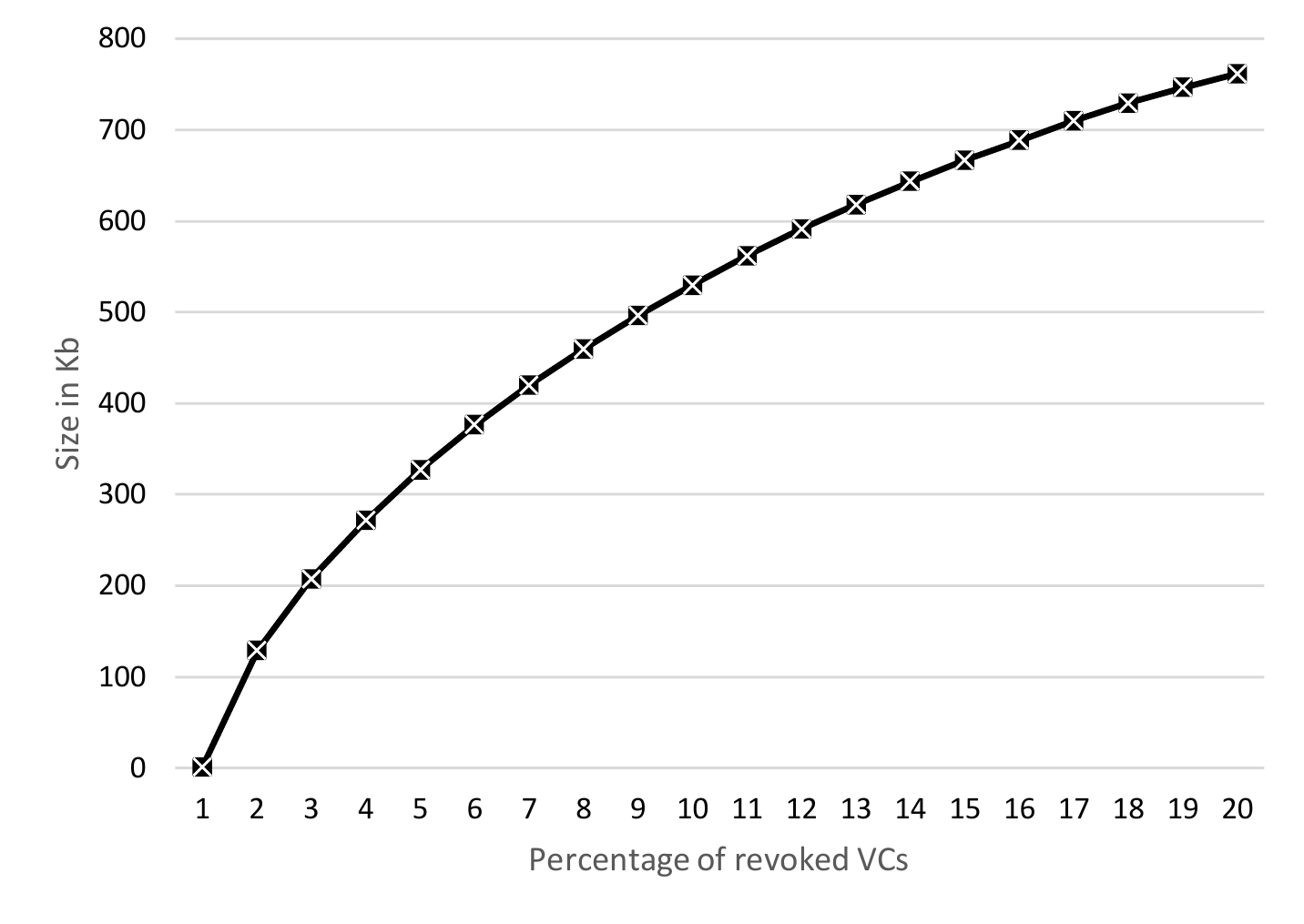}
\vspace{-0.5cm}
  \caption {Size of a VC revocation list
  % that includes 
  with
  1M VCs, as a function of the percentage of the revoked VCs.}
  \label{fig:rev_size}
\vspace{-0.5cm}
\end{figure}

\paragraph{Communication overhead} The communication overhead imposed to a PDP due to the usage control mechanisms of our solution depends on the number of PAPs that have issued the VCs included in the stored VPs. Particularly, if a PDP has stored the capabilities of $|C|$ consumers whose authorizations are provided by $|P|$ PAPs, the PDP has to send $|P|$ requests. Additionally, the size of the PAP response (i.e., the revocation list) depends on the number of VCs the PAP has issued (in general) and the percentage of the revoked VCs. Fig.~\ref{fig:rev_size} illustrates the size of a VC revocation list, provided by a single PAP, that includes $10^6$ VCs, as a function of the percentage of the revoked VCs. The figure shows that, because the list is compressed, a smaller 
% the 
number of 
%the 
revoked VCs yields a smaller list size. 

\subsubsection{Privacy}
%In our OAuth~2.0-based approach, a PAP learns when a data consumer interacts with an intermediary. This happens because  the PDP must receive a security token from the PAP in order for a consumer to get authorized. Therefore, the OAuth 2.0 approach facilitates tracking by the PAP. %On the other hand, since the consumer does not have to prove its identity to the Token endpoint, the evidence does not have to include any persistent consumer identifier (e.g., the real email address of a consumer). Similarly, since a  data intermediary does not have to maintain consumer specific information (e.g., it does not have to associate a consumer with an ``account''), every time a consumer is re-authorized an evidence may contain different personally identifiable information. Therefore, with the OAuth~2.0-based approach a consumer can be protected against data intermediary tracking (but not against PAP tracking).

%On the other hand, 
A consumer has to communicate only once with the PAP in order to obtain a VC; in all subsequent requests the PAP is not involved. Even during VC revocation status verification, a PAP cannot distinguish which specific VC a revocation status check concerns, since the revocation list includes information about all non-expired VCs.

%a VC contains a persistent consumer-specific identifier, which is used for proving VC proof-of-possession. As a result, whenever a consumer re-uses the same VC for getting authorized by a Token endpoint it can be tracked. Nevertheless, since the VC verification does not involve communication with the PAP, the PAP does not learn with which Token endpoint (and when) a consumer interacts. Even during VC revocation status verification, a PAP cannot distinguish which specific VC a revocation status check concerns, since a revocation list includes information about all non-expired VCs. Therefore, with the VC-based approach a consumer can be protected against PAP tracking (but not against data intermediary tracking). 

\subsubsection{Availability}
Our VC-based approach does not require any communication with the PAP during  consumer authorization. Although the PDP needs access to the revocation list, this does not have to be provided directly by the PAP. Since  the revocation list is included in a VC signed by the PAP, it can be provided by other entities or even the  consumer. In that case, the PDP should verify that the list is ``adequately fresh'' and that it has been signed by the appropriate PAP.

\section{Conclusions and Future Work}\label{conclusions}
In this paper, we presented an access control solution tailored for Data Spaces. Our approach is  Data Space API-aware and considers data semantics, enabling fine-grained policies. Furthermore, our Policy Enforcement Points are implemented as transparent HTTP proxies, making the protected resources oblivious to the access control mechanisms in place. This design ensures our solution can protect any Data Space system. We also presented a distributed approach that allows data owners to manage Policy Administration Points, enhancing security and increasing data owner sovereignty. %For this functionality we leverage both OAuth~2.0 and W3C Verifiable Credentials. OAuth~2.0 offers less complexity and enables continuous, fine-grained capabilities management, whereas Verifiable Credentials offer improved privacy and enable usage control with lower communication overhead. 

Our solution leverages iSHARE’s authorization model to implement capabilities-based access control. Future work will explore other access control paradigms, such as relation-based access control, and will integrate additional frameworks like the Open Policy Agent. We aim to extend our 
% Policy Enforcement Points 
PEPs
to apply access control policies to outgoing data, such as filtering based on user capabilities. Furthermore, our solution currently uses public keys to bind tokens and verifiable credentials to consumers; future research will investigate alternative identification methods, such as Decentralized Identifiers~\cite{did-primer}. Finally, we plan to enhance our solution to provide trustless
% Policy Enforcement Points 
PEPs
by encrypting data with keys bound to access control policies, such as Attribute-Based Access Control (ABAC).
%In this paper we presented an access control solution for data spaces.  Our solution is data space API aware and considers data semantics, enabling fine grain policies. Additional, our solution is modular and it allows modules related to access control administration to be administrated by data owners, improving this way the security of our approach and increasing the sovereignty of data owners. At the same time, our policy enforcement points are implemented as a transparent HTTP proxy, consequently the protected resource becomes oblivious to the deployed access control mechanisms; using this approach, our solution can be used for protecting any data space system.  

%Our solution implements capabilities-based access control by leveraging iSHARE's authorization model. Future work in this area includes the use of other types of access control paradigms (such as relation-based access control), as well as the integration of other access control frameworks (such as the open policy agent). To this end, our policy enforcement point can be extended to apply access control policies to outgoing data (e.g., filtering of data based on user capabilities). Similarly, our solution relies on public keys for binding tokens and verifiable credentials to consumers; future work in this area will investigate alternative means of  identification (e.g., Decentralized Identifiers~\cite{did-primer}). Finally, our solution can be extended to provide trustless policy enforcement points by encrypting data using keys bound to an access control policy (e.g., ABAC).  

% let's add it in the camera ready
%
 \section*{Acknowledgment}
% The work reported in this paper has been partly funded  by EU's Horizon 2020 Programme through the subgrant "A real-time AI-enabled worker safety preservation system" (MILESTONE) of project Trialsnet, under grant agreement No. 10109587.
This work has been funded in part
by EU's Horizon 2020 Programme, through the subgrant ``A real-time AI-enabled worker safety preservation system'' (MILESTONE) of project Trialsnet, under grant agreement No. 10109587.

\bibliographystyle{IEEEtran}
\bibliography{IEEEabrv,references}

\end{document}